\documentclass[a4paper,11pt]{article}
\usepackage{latexsym}
\usepackage[utf8]{inputenc}
\usepackage[activeacute,english,spanish]{babel}
\usepackage{lineno}
\RequirePackage{graphicx}
\RequirePackage{booktabs}

\usepackage[colorlinks,linkcolor={blue},citecolor={blue},urlcolor={red}]{hyperref}
\hypersetup{urlcolor=blue, colorlinks=true} 

\usepackage{geometry} 
\title{The Unified Era: An understanding journey from observations to the Unified Model of Active Galactic Nuclei}

\author{Leonardo de Lima 
 Santos\footnote{\url{https://orcid.org/0009-0006-8006-1998}} \\
 Postgraduate Program in Physics of the Instituto Tecnol\'ogico de Aeron\'autica (PG/FIS/ITA),\\ São José dos Campos, São Paulo, Brazil\\
\emph{leonardo.santos@pgfis.ita.br}\\
\and
Samuel Bueno Soltau\footnote{\url{https://orcid.org/0000-0002-7211-2533}} \\
 Physics Department, Institute of Exact Sciences, Federal University of Alfenas (UNIFAL-MG),\\ Alfenas, Minas Gerais, Brazil\\
 \emph{samuel.soltau@unifal-mg.edu.br}\\
}

\begin{document}
\maketitle

\selectlanguage{english}
\begin{abstract} \noindent
The Unified Model of Active Galactic Nuclei (UMAGN) is a comprehensive theoretical framework aimed at elucidating the diverse observations of AGN, encompassing quasars and Seyfert galaxies. The Model attributes the observed variations to different orientations of a surrounding matter disk around a supermassive black hole (SMBH), with the primary factor influencing observational diversity being the alignment of the AGN with the observer's line of sight. We present a comprehensive overview of the observational evidence, empirical and theoretical research, tracing key milestones that led to a unified perspective. We encapsulate the scientific journey culminating in the proposal of the UMAGN, including insights into the accretion disk, torus, and relativistic jet, and emphasize the properties of objects within UMAGN. Additionally, we underscore recent progress in multimessenger research involving electromagnetic waves, gravitational waves, astroparticles, and neutrinos, notably through collaborations such as the Event Horizon Telescope, LIGO/Virgo, IceCube, Pierre Auger, and KM3Net. We argue that these advancements present opportunities to enhance and refine UMAGN, contributing to a deeper understanding of AGNs and their implications for the formation and evolution of galaxies. The convergence of observational and theoretical research, coupled with emerging multimessenger techniques, paves the way for further strides in comprehending these enigmatic cosmic phenomena.\\
\textbf{Keywords}: Astrophysics, Unified Model of Active Galactic Nuclei, Black Holes.
\end{abstract}

\newpage

\selectlanguage{spanish}
\begin{center}
\textbf{La Era Unificada: Una trayectoria de comprensión desde las observaciones hasta el Modelo Unificado de Núcleos Galácticos Activos}
\end{center}

\begin{abstract} \noindent
El Modelo Unificado de N\'ucleos Activos Gal\'acticos (MUNAG) es un marco te\'orico destinado a esclarecer las diversas observaciones de N\'ucleos Activos Gal\'acticos (NAG), que abarcan cu\'asares y galaxias Seyfert. El modelo atribuye las variaciones observadas a diferentes orientaciones de un disco de materia circundante alrededor de un agujero negro supermasivo, siendo el factor principal que influye en la diversidad observacional la alineaci\'on del NAG con la l\'inea de visi\'on del observador. Presentamos una visi\'on general de la evidencia observacional, la investigaci\'on emp\'irica y te\'orica, rastreando hitos clave que llevaron a una perspectiva unificada. Resumimos el viaje cient\'ifico que culmin\'o en la propuesta del UMAGN, incluyendo percepciones sobre el disco de acreci\'on, el toro y el chorro relativista, y destacamos las propiedades de los objetos dentro del UMAGN. Adem\'as, subrayamos los avances recientes en la investigaci\'on multimensajera que involucra ondas electromagn\'eticas, ondas gravitacionales, astropart\'iculas y neutrinos, notablemente a trav\'es de colaboraciones como el Telescopio del Horizonte de Eventos, LIGO/Virgo, IceCube, Pierre Auger y KM3Net. Sostenemos que estos avances ofrecen oportunidades para mejorar y refinar el UMAGN, contribuyendo a una comprensi\'on m\'as profunda de los NAG y sus implicaciones para la formaci\'on y evoluci\'on de las galaxias. La convergencia de la investigaci\'on observacional y te\'orica, junto con las emergentes t\'ecnicas multimensajeras, allana el camino para nuevos avances en la comprensi\'on de estos enigm\'aticos fen\'omenos c\'osmicos.\\
\textbf{Palabras clave}: Astrofísica, Modelo de Núcleos Galácticos Activos, Agujeros Negros.
\end{abstract}

\section{Introducción}

The term ``AGN'' stands for Active Galactic Nucleus, which refers to the compact region at the center of a galaxy that are characterized by their significant and unusual activity, such as the emission of large amounts of electromagnetic radiation, including X-rays and radio waves. It is believed that this high activity is due to the presence of a supermassive black hole (SMBH) that is accreting mass from its surroundings~\cite{Peterson1997,Martinez2006}.
Nowadays, it is widely accepted that the central engine of the AGN is a SMBH, an idea originally proposed in the mid-1960s by Salpeter~\cite{Salpeter1964}, Zel'dovich and Novikov~\cite{Zeldovich1964}, and Lynden-Bell~\cite{Lynden-Bell1969}.

For decades, since the Seyfert's observations, in the 1940s~\cite{Seyfert1943}, through Schmidt's 3C 273 radio source in 1963~\cite{Schmidt1963}, the number of apparently distinct objects has increased, with only their strangeness and unusual spectral behavior in common.
It is reasonable to assume that with the gradual technological advancement of instruments, the improvement of astronomical observation techniques and the increase in research, the diversity of nomenclatures (Seyfert, quasar, blazar, BL Lacertae, radio galaxies, etc.) that apparently designated different astrophysical objects led to the advent of the ``Unified Era'' in the mid to late 1980s. So, despite the apparent dissimilarity among AGN, a synthesis of empirical evidence gathered from systematic observations and theoretical studies emerged in the 1980s. Antonucci consolidated the entire known range of AGN features known at the time into one model during this period, marking the inception of the Unified Era and the Unified Model of Active Galactic Nuclei (UMAGN)~\cite{AntonucciMiller1985}. The UMAGN posits the presence of a supermassive black hole (SMBH) at the center, surrounded by an accretion disk and a torus, and two relativistic jets emitted perpendicular to the plane of the accretion disk. The diversity of objects observed from a terrestrial point of view is elucidated by varying the perspective of the components comprising UMAGN.

In this study, our objective is to demonstrate that the literature describing objects such as quasars, Seyfert galaxies, BL Lacertae, and other active galactic nuclei (AGNs) provides the evidence that has accumulated and enabled the emergence of the synthesis proposed by~\cite{AntonucciMiller1985, Antonucci1993, Antonucci2012} in the Unified Era. Conversely, we acknowledge that articles in the post-Unified Era have played a role in consolidating and refining the Unified Model of AGNs (UMAGN). Finally, we argue that in the contemporary moment, the responsibility for providing evidence that corroborates or improves UMAGN lies with collaborations such as the Event Horizon Telescope (EHT), LIGO/Virgo, IceCube, Pierre Auger and KM3Net, which offer multimessenger information and contribute to the ongoing research on AGNs.

The structure of this article is as follows.
In Section~\ref{sec:umagn}, we describe some of the characteristics and physical properties of AGNs that provide the variety of phenomena that created the apparent variety of observed objects and how they were explained in the UMAGN synthesis.
In Section~\ref{sec:unifiedera}, we present an overview, without claiming to be complete, of the evidence, discoveries and theoretical research that emerged in the Era of Unification and that culminated in the elaboration of UMAGN.
In Section~\ref{sec:possibilities}, we explore the possibilities offered by recent experiments carried out by the Collaborations and how they could contribute to improving the UMAGN and advancing our knowledge about AGNs with the contribution of multi-messenger data and information.
Finally, in Section~\ref{sec:conclusion}, we summarize the presented perspective and discuss its implications for our understanding of the evolution of AGNs and the black holes at their centers.

\section{The Unified Model of Active Galactic Nuclei}
\label{sec:umagn}

The UMAGN developed by Antonucci \& Miller~\cite{AntonucciMiller1985} by the end of the Unified Era, in the mid-1980s, offers an explanation the diverse observed properties of AGNs as being produced by orientation-dependent effects of a relativistic jet, an optically thick torus and an accretion disk  surrounding a central SMBH. AGNs are into central regions of galaxies, where a SMBH accretes matter from its surroundings, leading to the release of large amounts of energy. The radiative processes and energy regimes that produce the registered emissions from AGNs include accretion disk electromagnetic emission, broad and narrow emission lines, and non-thermal continuum radiation, such as synchrotron and inverse Compton emission.

The typical effects observed in captured data signatures from Earth include a wide range of electromagnetic radiation across the spectrum, from radio waves, X-rays to gamma rays. These emissions are often characterized by strong variability on various timescales, suggesting the presence of compact emission regions. The observed signatures also include the presence of relativistic jets, which can produce intense outbursts and flares in the optical, radio and gamma-ray bands. These phenomena provide information about the energetic regimes and physical processes taking place within the innermost regions of AGNs, which are in some cases beyond the resolution of current instruments.
These physical processes taking place within AGNs, such as the emergence of flares due to the injection of high-energy particles by shock waves passing along the relativistic jets, as well as the role of magnetohydrodynamical instabilities near the base of the jets in producing energetic events. 

\begin{figure}
\centerline{\includegraphics[width=0.89\textwidth]{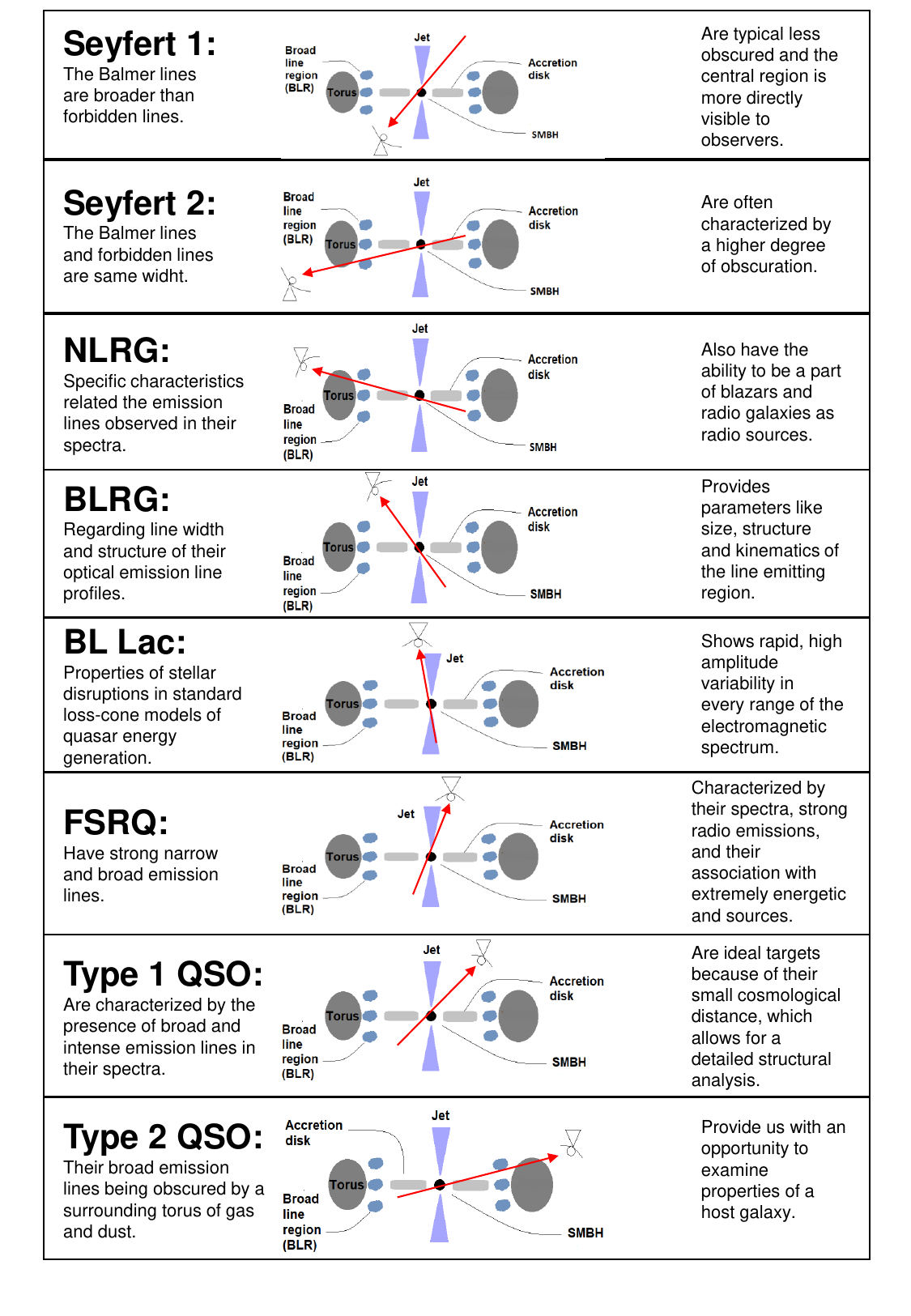}}
\caption{Variety of objects explained by UMAGN ans some characteristics.}
\label{fig:umagn}
\end{figure}

The specific types of AGNs and their differences in terms of physical phenomena and signatures are explained in the UMAGN  encompass various types, including quasars, blazars, BL Lacertae and Seyfert galaxies. These types differ in terms of their observed characteristics and physical properties (see Fig.~\ref{fig:umagn}). 
The differences in physical phenomena and data signatures among these AGN types arise from variations in their accretion rates, orientation, and the presence of relativistic jets. For instance, blazars exhibit rapid and large-amplitude flares in the optical, X-ray and gamma-ray bands, which are attributed to magnetohydrodynamical instabilities near the base of the jets. Quasars, on the other hand, are known for their strong and broad emission lines, indicating the presence of highly ionized gas in their vicinity, are highly luminous and powered by accretion onto SMBH, while blazars are characterized by relativistic jets pointing toward Earth, leading to intense variability observed across the electromagnetic spectrum. Seyfert galaxies, on the other hand, have less luminous nuclei.

\subsection{Interpreting signatures received on Earth}\label{sec:interpr}

In the study of AGNs radiative processes, distinct regions within these structures offer insights into the astrophysical phenomena occurring around SMBH~\cite{MoBosh2010}. As we delve into the intricacies of each region, we uncover the origins and radiative processes that contribute to the observed spectral signatures.

Beginning with the Broad-Line Region (BLR) situated at the core, emissions emanate from the proximity of the supermassive black hole. Notably, broad spectral lines such as $H\alpha$, $H\beta$, and Mg II arise from the recombination of ionized atoms, providing a unique glimpse into the conditions near the central nucleus.

Moving outward, the accretion disk plays a pivotal role in AGN dynamics. Material undergoing accretion toward the supermassive black hole generates thermal emissions, prominently featuring X-ray radiation from the hot disk. This Accretion Disk's radiative processes contribute significantly to the energetic output of AGNs.

In the periphery lies the Narrow-Line Region (NLR), an expanse beyond the accretion disk. Emissions in this region manifest as narrow spectral lines, such as [O III] and [N II], resulting from ionization induced by ultraviolet radiation originating from the central disk. Understanding the NLR provides valuable insights into the extended regions surrounding AGNs.

Examining the energetic regimes and observational effects, we find that X-ray emissions predominantly emanate from the Accretion Disk, a phenomenon detectable by observatories like Chandra and XMM-Newton. Broad-Line Regions, influenced by ionizing radiation from the central nucleus, exhibit broad spectral lines observable in optical and ultraviolet spectra.

Conversely, the Narrow-Line Region, ionized by radiation from the central disk, yields narrow spectral lines, including [O III] and [N II], detectable in optical spectra. Additionally, optical polarization arises from the scattering of light in the Molecular Torus, introducing a variable optical polarization contingent on the observer's viewing angle.

\section{The Unified Era}
\label{sec:unifiedera}

Antonucci and Miller~\cite{AntonucciMiller1985} defines the so-called Unified Era as a diversity in AGN types, that can be attributed to varying orientations relative to the line of sight. One extreme hypothesis is the straw person model (SPM), which suggests that there are two fundamental types of AGN: radio quiets and radio louds. While this model is a simplified representation of the unification idea, it has been mostly ruled out on various grounds. The discussion revolves around convincing evidence that orientation effects are both important and widespread. The orientation of an object relative to the observer can significantly affect its classification and properties. The true situation likely falls somewhere between the SPM and the idea that orientation doesn't influence classification at all. Your conclusion suggests that the SPM represents significant progress in understanding orientation effects in AGN and quasars. While the strict SPM is not entirely accurate and may not fully describe the situation, it has laid the foundation for further refinement of our understanding of orientation-dependent classifications.

In 1943, Seyfert~\cite{Seyfert1943} observed the emission lines in the spectra of extra galactic nebulae, with a focus on a rare class of objects characterized by high-excitation emission lines superposed on typical spectra. Mostly intermediate type spirals with luminous or semi-stellar nuclei containing a significant portion of the total light, belong to this unusual category. Seyfert references early observations of these objects, that exhibit emission lines with significant width, and the highlights observation of the emission lines being small discs or bands several angstroms wide.

In 1963, Schmidt~\cite{Schmidt1963} discussed observations of an object that show broad emission features on a blue continuum. The presence of certain emission lines suggests a Redshift, indicating a high apparent velocity. There are two possibilities as explanations: the object is either a star with a large gravitational Redshift or the nuclear region of a distant galaxy wit a cosmological Redshift. Schmidt explained that is favored due to the observed properties.

In 1963, Matthews~\cite{Matthews1963} explained about the position accuracy that has greatly improved the efficiency of the search for optical identifications compared to earlier searches years ago. These identifications revealed that radio sources are associated with a variety of astronomical objects. The distribution of discrete radio sources above a certain declination has been found to be isotropic and is generally attributed to galaxies. Before these identifications, no star, had been associated with a radio source, except for the Sun.

In 1969, Lynden-Bell~\cite{Lynden-Bell1969} debated the idea that Quasars evolve into powerful radio sources with two well-separated radio components. The energies involved in these outbursts are calculated to be enormous, and it is suggested that gravity may play a dominant role in providing the necessary energy. Discusses various astronomical observations related to different galaxies and their central regions. Lynden-Bell mentions observations in other galaxies, and provides estimates for nuclear masses and fluxes. Discusses Seyfert galaxies and their activity levels, a measure of flux, is high active Seyfert galaxies due to the presence of a significant amount of gas in their central regions. Explained the rapid buildup of mass is the Schwarzschild throat large values, suggests that during the formation of galaxies, there was a significant amount of gaseous material present. Also references recent observations of the galactic center and propose a dust model to explain infrared observations  Discusses various astronomical observations related to different galaxies and their central regions.

In 1977, Blandford and Rees~\cite{Blandford1977} focused on the interpretative aspects of observations of strong radio sources in galactic nuclei. Provided insights and interpretations into the historical significance of strong radio sources as evidence of violent activity in galactic nuclei Jets are indicative of some type of asymmetry in the collimation mechanism. There are two possibilities for jets: The emission can come from particles accelerated by dissipation processes; or the electrons in the beam itself may not have been completely cooled by radiative or adiabatic losses during exit from the nucleus. If it is the former, it would predict brightening of the limbs and a magnetic field along the jet; if it is the second case, it would predict a magnetic field perpendicular to the jet.

Until now, all articles have contributed to the creation of the unified AGN model. At this moment, the unified era will be presented and subsequently the work that culminates in its improvement.

\subsection{The Unified Era}

In 1982, Antonucci~\cite{Antonucci1982} studied the alignment of optical polarization position angles with the large-scale radio structure in low-polarization quasars. Such alignment implies a geometrical relationship between the inner, optically-emitting region and outermost, radio-emitting region. He still argues two possible causes for optical polarization: synchrotron emission and scattering. If polarization is due to synchrotron emission, it reveals the orientation of the magnetic field in the optically-emitting region. If scattering is the cause, the position angle reflects the distribution of scatterers. Antonucci mentions their observations of quasars in a sample and discusses a forthcoming study to distinguish between these possibilities. Emission-line polarizations suggest that polarizations in both groups are likely due to scattering or dust transmission rather than the emission process itself. Indicates their intention to discuss the implications of these results in more detail in a different context, where they will have access to the complete set of radio, optical spectroscopic, and optical spectropolarimetric data.

In 1991, Roche~\cite{Roche1991} studied the infrared (IR) energy distributions of various classes of galaxies by combining photometric data from IRAS (Infrared Astronomical Satellite) mission with ground-bases observations at shorter wavelengths. The most luminous AGN with powerful non-thermal emissions tend to have IR SEDs that smoothly connect with optical and radio observations, implying continuity in the emission mechanism. In contrast, lower-luminosity AGN show evidence of an additional IR component, often peaking in the infrared, which is best explained as emission from dust grains. Galaxies with active nuclear star formation also exhibit strong thermal peaks in the infrared. Even normal spiral galaxies emit significantly in the IR, with a substantial portion of their output originating from their galactic discs. In contrast, active nuclei often exhibit less structure spectra, which can often be approximated by power-law distributions at these wavelengths.

In 1995, Urry~\cite{Urry1995} delves an enigmatic nature of AGN and their significance in understanding the universe. AGN are unique in that they generate extremely high luminosities in a concentrated volume, likely through processes other than nuclear fusion, which powers stars.  The prevailing model of AGN's physical structure involves a supermassive black hole at the center, the gravitational potential energy of which is the source of AGN luminosity. Matter spiraling into the black hole emits radiation in the form of an accretion disk, primarily in the ultraviolet and soft X-ray wavelengths. The presence of strong optical and ultraviolet emission lines is attributed to rapidly moving clouds of gas in the gravitational potential of the black hole, known as broad-line clouds. These emissions, are obscured along certain lines of sight by a torus or warped disk of gas and dust. Energetic particles are expelled along the poles pf the disk or torus, forming collimated radio-emitting jets and sometimes giant radio sources. This inherently asymmetric model implies that AGN appear radically different at different viewing angles. To reconcile these differences in appearance, AGN of various orientations are assigned to different classes, and unification is essential to study the underlying physical properties.

\section{P\'os Unified Era}


In 2000, Elvis~\cite{Elvis2000} proposed a simple yet comprehensive structure to explain the inner regions of quasars, particularly focusing on their emission and absorption line phenomena. This model successfully accounts for both broad and narrow absorption lines and also provides explanations for other emission line and scattering effects. Quasars are a solvable problem, some coherent structure must be present. Quasars research results in a single simple scheme: he high-and-low-ionization parts of the BELR,the BAL and NAL regions, and the Compton-thick scattering regions can all be combined into the single funnel-shaped out-ow. All of these features come about simply by requiring a geometry and kinematics constructed only to explain the two types of absorption lines. This unification gives the model a certain appeal.

In 2001, Vollmer~\cite{Vollmer2001} provided information about the structure and characteristics of the Circumnuclear Disk (CND) surrounding the Galactic Center. The CND, consists of gas and dust clouds and was initially discovered as a tilted dust ring. It has a clumpy structure, low ionization, and is influenced by the radiation from the central Heistar cluster. A continuous, smooth molecular disk model is ruled out due to its clumpy nature. The CND rotates, has a sharply defined inner edge, and is inclined to the line of sight  Presents an analytical model for the Galactic Center's Circumnuclear Disk, that consists of around 500 heavy clouds, creating a disk-like structure. The model successfully matches observations and identifies two stable cloud regimes: heavy molecular clouds and their stripped cores.

\cite{Kormendy2013}

\section{Collaborations \& Multimessenger Astronomy}
\label{sec:possibilities}

Although we already have evidence of the existence of SMBHs in some cases, including in the Milky Way~\cite{Ghez2008}, and an image of Sagittarius $A^{\ast}$~\cite{EventHorizonTelescope2019} and practically all other alternatives to SMBHs are currently ruled out, and Occam's razor strongly suggests that many, if not all, galaxies do in fact host SMBHs that can fuel AGN activity.

However, there are still questions that remain to be understood~\cite{MoBosh2010}. Even though the idea of unification is widely accepted, there are still many open questions that need to be understood. Some of the basic ideas related to the unification of the AGN require observational evidence as they are not yet conclusive. AGN formation can be understood within the framework of galaxy formation, but the two most important conditions for the production of an AGN are the existence of a central SMBH and a sufficient amount of gas to fuel the nucleus. Understanding how SMBHs form and the mechanisms responsible for transporting gas towards the center of the host galaxy to feed the black hole are key aspects that need to be further explored.

An important fact that any theory of AGN formation must take into account is that quasars are observed at high redshifts. The time scale for the formation of an SMBH must be less than cosmic time at these redshifts. The growth of SMBHs and the promotion of AGN are areas that require further investigation to fully understand the processes involved.

The kinematics of stars and gas in the central regions of spheroidal galaxies indicate the presence of massive dark objects, but further investigation is needed to determine whether these objects are SMBHs or other alternatives, such as dense clusters of stellar remnants or exotic objects such as fermion balls. heavy stars or bosons. Probing kinematics at smaller scales and detecting higher velocities will provide more information about the nature of these central massive objects. For this reason, the contribution of multimessager data from Collaborations is so essential.

The quest to unravel the mysteries of AGNs and the SMBHs residing at their cores has entered a new era with the advent of multimessenger astronomy. This transformative approach relies on combining information from different cosmic messengers, such as electromagnetic waves, gravitational waves, and neutrinos, to gain a comprehensive understanding of the most extreme astrophysical phenomena. One of the pioneering endeavors in this field is the Event Horizon Telescope~\cite{EHTweb2024} (EHT), a global collaboration that links radio telescopes across the planet to form a virtual Earth-sized telescope. The EHT's unprecedented resolution allows for the observation of the immediate vicinity of SMBHs, providing vital clues about their accretion processes and the dynamics of the surrounding regions.

Gravitational wave astronomy has emerged as another indispensable tool for probing the universe's most enigmatic phenomena, including AGNs and SMBHs. The Laser Interferometer Gravitational-Wave Observatory~\cite{LIGOweb2024} (LIGO) and Virgo collaborations have significantly expanded our ability to detect gravitational waves, enabling the observation of cataclysmic events such as the merger of black holes. By combining gravitational wave signals with electromagnetic observations, we can glean insights into the formation, growth, and merger history of SMBHs. The synergy between EHT and LIGO/Virgo is particularly potent, as it offers a unique opportunity to study AGNs and their associated gravitational wave signatures simultaneously, opening new avenues for understanding the intricate interplay between matter, space, and time.

In the realm of neutrino astronomy, experiments like IceCube~\cite{IceCubeweb2024} have become instrumental in providing complementary information about the high-energy processes occurring in AGNs and the vicinity of SMBHs. Neutrinos, being elusive and nearly massless particles, can traverse vast cosmic distances without interaction, carrying information about the extreme environments where they originated. By combining IceCube's neutrino data with observations from other instruments, we can create a more comprehensive picture of the energetic processes associated with AGNs and gain insights into the nature of their central black holes.

Furthermore, the Pierre Auger Observatory~\cite{PierreAugerweb2024} and the upcoming KM3NeT~\cite{KM3NeTweb2024} project contribute crucial components to the multimessenger puzzle. Pierre Auger's focus on ultra-high-energy cosmic rays complements the information provided by other messengers, shedding light on the cosmic accelerators responsible for producing such particles in the vicinity of AGNs. Meanwhile, KM3NeT, designed to detect high-energy neutrinos in the deep sea, adds another layer to the multimessenger approach. The synergy between these experiments allows scientists to cross-validate and cross-correlate information from different messengers, enhancing the reliability and completeness of our understanding of AGNs and the SMBHs embedded within them.

The collaborative efforts of projects like the Event Horizon Telescope, LIGO/Virgo, IceCube, Pierre Auger, and KM3NeT are essential for advancing our knowledge of AGNs and the SMBHs residing at their cores. The combination of electromagnetic, gravitational wave, and neutrino observations provides a holistic view of these extreme astrophysical environments, allowing us to address fundamental questions about the formation, evolution, and behavior of AGNs and their central black holes. As we delve deeper into the multimessenger era, the synergy between these diverse observational techniques will undoubtedly propel us towards unprecedented discoveries and a more profound understanding of the cosmic phenomena that shape our universe.

\section{Conclusion}
\label{sec:conclusion}

In conclusion, the Unification Era represents a significant advancement in our understanding of AGNs, connecting the previously separate fields of radio-loud and optically-luminous AGNs and providing a more complete picture of these enigmatic objects.

\section*{Acknowledgments}

We would like to thank Dr. Ricardo Bulcão Valente Ferrari and Dr. C\'assius Anderson Miquele de Melo for helpful comments and discussions. This study was financed in part by the Coordenação de Aperfeiçoamento de Pessoal de Nível Superior -- Brasil (CAPES) -- Finance Code 001.

\bibliographystyle{unsrt}
\bibliography{main}

\end{document}